\documentclass[twocolumn,showpacs,preprintnumbers,amsmath,amssymb,floatfix]{revtex4}

\usepackage{graphicx}
\usepackage{dcolumn}
\usepackage{bm}

\begin{document}

\preprint{APS/123-QED}

\title{The prisoners dilemma on a stochastic non-growth network evolution model}
\author{Vasilis Hatzopoulos}
\author{Henrik Jeldtoft Jensen}
\email[Author to whom correspondence should be addressed:\\]
{h.jensen@imperial.ac.uk} 
\homepage{http://www.ma.imperial.ac.uk/~hjjens/}
\affiliation{Department of Mathematics, Imperial College, 
180 Queen's Gate, London SW7 2AZ, U.K.}

\begin{abstract}
We investigate the evolution of cooperation on a non - growth network model with death/birth dynamics. Nodes reproduce under selection for higher payoffs in a prisoners dilemma game played between network neighbours. The mean field characteristics of the model are explored and an attempt is made to understand the size dependent behaviour of the model in terms of fluctuations in the strategy densities. We also briefly comment on the role of strategy mutation in regulating the strategy densties.
\end{abstract}
\maketitle
\section{Introduction}

In biology an organism is called an altruist, if its behaviour enhances the reproductive probability (fitness) of another individual at the expense of its own. An altruist then pays a cost $c$ for its opponent to receive a benefit $b$ while a non altruist pays nothing and just receives the benefit. Should such behaviours be genetically encoded and if natural selection is solely interpreted as preserving the traits that contribute to individual fitness, then altruistic behaviours should be at a selective disadvantage with repsect to purely selfish ones. Darwin himself was aware of this paradox which he saw in the proliferation of sterile insect castes. 

A mathematical integration of altruism into evolutionary theory  did not appear until the 1960's when Hamilton proposed that under certain conditions altruism can be favoured by natural selection if the recipient and donor of the altruistic act are genetically related ~\cite{Hamilton} a mechanism coined kin selection. Genetically related individuals have more common alleles  than two individuals picked at random, and whilst the donor of the altruistic act lowers its  reproductive chances by doing so it raises the fitness of its relative who has a good chance of also carrying the allele coding for  altruistism. Thus Hamilton posited that selection can also act as to increase an organism's inclusive fitness,  comprised of the reproductive potential of an individual plus its genetic relatives, and the family can be seen as the seed from which altruistic behaviour grows. Mathematically Hamiltons rule takes the form $\frac{b}{c} > \frac{1}{r}$, where $c$ is the cost of altruism to the actor, $b$ is the benefit to the recipient  and $r$ is a coefficient measuring the relatedness of the two. An important conceptual shift included in this viewpoint is that selection acts not as to preserve individual organisms but genes in a pool spanning whole populations. Trivers ~\cite{Trivers} proposed that altruism can evolve if it is reciprocated even if the individuals involved are not related. Group selection ~\cite{SoberWilson} holds that altruistic behaviour can evolve if selection can act at the level of the group, i.e. a population with more altruists will reproduce faster than a population with fewer altruists. The idea remains controversial to this day mainly due to the fact that altruistic groups are susceptible to the invasion of cheaters.

Game theory, particularly in its evolutionary form first introduced in ~\cite{SmithPrice} and ~\cite{MSmith}, has proved to be a very useful tool in studying the evolution of behaviour because it fundamentally describes frequency dependent selection. The worth of a particular behaviour with regards to selection depends on its frequency in the population and thus the behaviour of others, an altruistic behaviour can only be deemed adaptive if there is a sufficient frequency of altruists.
The prisoner's dilemma game ~\cite{AxelrodHamilton} has been extensively used to model the evolution of cooperation/altruism between non-related individuals.
The game describes an interaction between two players each having two strategies available to them, to cooperate, $c$ or to defect $d$. A $c - c$ outcome yields the payoff $R$ for both players while a $d - d$ event will give both players  a payoff $P$. Should one player cooperate and the other defect the cooperator will collect $S$ and the defector $T$. The payoffs are arranged such that $ T >R>P>S$ and $2R>T+S$. In this setting it is always better to defect irrespective of the other player's choice . The dilemma lies in that if both players defect they both receive $P$ instead of $R$ should they have both cooperated. In the evolutionary game theory setting strategies become phenotypes and payoffs become fitness. Games take place between members of the population in a pairwise fashion and the fittest phenotypes spread over the population.

In the 1980s Robert Axelrod organized tournaments ~\cite{AxelrodBook} whereby contestants were requested to submit strategies to play the repeated prisoners dilemma  against each other for a finite but unknown number of times. The strategies were recipes that specified the players next move based on stored knowledge of the opponents past moves. Strategies would then reproduce themselves in proportion to their accumulated payoff after having played with a representative sample of the population. In this setting it was demonstrated that cooperation is an evolutionary stable outcome through a process of reciprocity as evident by the proliferation of the well known tit-for-tat strategy. Tit-for-tat will coopearate until defected against and will then keep defecting until the opponent cooperates at which point it will switch back to cooperation. Reciprocity requires some sort of opponent recognition mechanism (and thus advanced cognitive faculties) which was deemed sufficient but not necessary for the evolution of cooperation. 

The need for cognitive capacity was done away with in ~\cite{NowakMay1992} were pure $c$ or $d$ strategies occupied the points of a two dimensional grid. The strategies  would play their grid neighbours tally their total score and then for the next round copy the strategy of their highest scoring neighbour. Cooperative elements could now survive on the grid by forming clusters uninvadable by defecting elements at their border. The need for memory was thus replaced with spatial structure, much more prevalent in the natural world then advanced cognitive capacity. Variations in the lattice connectivity, topology and strategy update mechanism have since been extensively studied (~\cite{Lindgren1994}, ~\cite{SzaboToke}, ~\cite{KillingbackDoebelli}, ~\cite{VanBaalen}, ~\cite{Vainstein}, ~\cite{AbKup2001}, ~\cite{EbelBornholdt}, ~\cite{JunKimEtAl}, ~\cite{Xiaojie},~\cite{HolmeEtAl}, ~\cite{Vukov}, ,~\cite{Zimm2005}, ~\cite{Santos}, ~\cite{TaylorDay}). Doing away with the need for an update mechanism (and thus any semblance of cognitive ability in the players/agents) and also introducing stochastic dynamics the authors in ~\cite{NowakReplicator} and ~\cite{OhtsukiNowakB}  studied different lattice topologies for pure strategy populations evolving under the Moran process. This two step process consists of a death event and a birth event (which can be performed in any order). The death event involves the removal of a randomly selected element while the birth event occurs when the now vacant vertex gets colonized by a copy of the highest scoring neighbour.

In this work we also employ a two step death/birth process with the following differences. When an individual is removed all its associated edges are also deleted from the system. When an individual is born, as a copy of a stochastically chosen fit individual, links are created between the newborn and other individuals through a process controlled by three parameters.
Thus the individuals dynamically create the network which is not a pre-existing space to be colonized but an ever changing relational web. Also after the death of a node any fit node and not only its neighbours can parent an offspring which implies that clustering is not necessarily the mechanism for the evolution of cooperation in our work.

\section{Network dynamics and selection}

Before we introduce our algorithm consider a population of size $n = n_{c}+n_{d}$ with the strategies present in relative abundances $\rho_{c}=\frac{n_{C}}{n}$ and $\rho_{d}=\frac{n_{D}}{n}$. In the well mixed scenario every pair of interactions in the population is present, which permits us to write
\begin{equation}W_{c} = \rho_{c}R + \rho_{d}S \end{equation}
\begin{equation}W_{c} = \rho_{c}T + \rho_{d}P\end{equation}
\begin{equation}
\overline{W}=\rho_{c}W_{c}+\rho_{d}W_{d}
\end{equation}
for the fitness of a cooperator, defector and the mean fitness respectively.

In the limit $n\rightarrow\infty$ the evolutionary game dynamics of this system are well described by the replicator equation~\cite{TaylorJonker}. The replicator is a deterministic equation that, given a strategy composition, calculates the composition in the next generation. The replicator assumes that strategies reproduce at a rate proportional to their fitness.
In its discrete form this equation reads:
\begin{equation}
\rho_{c}^{'}=\frac{\rho_{c}W_{c}}{\overline{W}} 
\label{discReplicator}
\end{equation}
with a similar expression existing for the defector population. To compute then the evolutionary history of the population we repeatedly calculate the fitnesses and then use the replicator equation to create the next generation up until the point where $\rho_{c}^{'}=\rho_{c}$. At this point, and in the absence of mutation, there is no further evolution and the population is said to be in an evolutionary stable state(ESS). The ESS may be monomorphic(one strategy present) or polymorphic (coexisting populations). If it is a polymorphic ESS then the different strategies present in the population should be at fitness equilibrium. Since the prisoners dilemma is defined by the relations $T>R>P>S$ we have that $W_{d} > W_{c}$ and a population evolving under natural selection with no mutation will eventually reach the monomorphic state $n_{c}=0, n_{d}=1$.

In large networks of actors all possible pairs of interactions are rarely present, so now consider a process that in every subsequent round of the repeated game matches each player with a randomly selected subset of the population. Then in the mean filed limit $W_{c} = k\rho_{c}R + k\rho_{d}S$ and $W_{d} = k\rho_{c}T + k\rho_{d}P$, where $k$ is the average number of interactions per player. Repeated application of the replicator equation will then again yield a population of $n_{d}=n$ in this unstructured population. The question arises then whether the existence of a polymorphic ESS is possible in the prisoner's dilemma and what are the mechanisms that can bring it forth.

Here we maintain and evolve a population of pure strategists linked by a network whose degree distribution evolves under an algorithm first introduced in ~\cite{LairdJensen}. In its original formulation a time step of the algotithm consists of a stochastic death/birth proccess.
\begin{itemize}
\item Removal. Choose a node at random, with probability $\frac{1}{n}$, and remove together with all its associated edges
\item Duplication. Choose a node at random, with probability $\frac{1}{n-1}$, this is the parent node. Introduce a new offspring node and attach edges between the node and the remaining $n-1$ nodes in the following way.
\begin{enumerate}
\item Attach the offspring to the parent with wiring probability $p_{p}$.
\item Attach the offspring to other non-parent nodes with wiring probability  $p_{o}$ if an edge does not exist between the parent and the other node.
\item Attach the offspring to other non-parent nodes with wiring probability  $p_{e}$ if an edge exists between the parent and the other node.
\end{enumerate}
\end{itemize}

To the above algorithm we add selection, based on the payoffs a node collects by playing against its neighbourhood, and also let offspring inherit the strategy from  their parent. In our simulations we let the population evolve under tournament selection ~\cite{Blickle}.In tournament selection a subset of the population is selected at random and all selected play against the nodes in their respective neighbourhoods. Their cumulative payoffs are compared and the node with the highest score parents a new offspring. In our simulations the subset consists of two randomly picked nodes, if they happen to have the same total payoff we randomly select one to reproduce. While the replicator Eq.(~\ref{discReplicator}) is generational our tournament selection only replaces a randomly selected individual with the winner of the tournament. This means that one application of the replicator should correspond on average to $n=pop.size$ death/birth cycles.


To initiate our system we cast a random network over the population such that $k=p_{o}n$ is the mean degree. We also randomly and with equal probability assign a  $c$ or $d$ strategy to each individual. Concerning the payoffs we also let $T>R>P=S=0$ so that only cooperative links have a positive effect on fitness, a choice that does not alter the essential characteristics of the prisoner's dilemma (~\cite{NowakMay1992}).

If we set $p_{o}=p_{e}=0, p_{p}=1$ nodes will only interact with their same strategy parent and the population will settle in a $n_{c}=n$ state since $R>P$. On the other hand letting $p_{p}=p_{e}=0$ and $0<p_{0}\leq1$ the only stable equilibrium is the $n_{d}=n$ state. Under our network dynamics it is possible for a node to become isolated even if all the wiring probabilities are greater than zero. This occurs if a node has all of its neighbours progressively removed through its lifetime by death events. An isolated node has no payoff  and selection is blind to its strategy since the phenotype only expresses itself through the inclussion of a node to a neighbourhood. Such a node can still reproduce by chance (for example by being compared in the tournament against another node only connected to defectors)and re-attach itself to the network through its offspring.

In this set of experiments we examine the parameter set $0<p_{o}<<1, p_{e}=0, p_{p}=1$ which describes a world where elements link up with their parent and a randomly chosen subset of the population which we vary in size. In this limit a node will on average end up with $p_{o}n$ randomly formed connections plus a link to its parent. The node average degree will then be $k=p_{o}n+1$.
The population structure introduced by the wiring probabilities can then be expressed in  the mean fitness calculations as(excluding self interaction):
\begin{equation}
\overline{W_{c}}=(k-1)\frac{n_{c}-1}{n-1}R+R
\label{wc}
\end{equation}
for a cooperator and:

\begin{equation}
\overline{W_{d}}=(k-1)\frac{n_{c}}{n-1}T
\label{wd}
\end{equation}
for a defector.
Under natural selection for a polymorphic population to exist the expected fitnesses of the two types must be equal. Setting $\overline{W_{c}}=\overline{W_{d}}$ we can calculate the equilibrium cooperator density to be:
\begin{equation}
\rho_{C}=\frac{n-k}{n}\frac{R}{(T-R)}(k-1)^{-1}
\label{systemDepEquil}
\end{equation}
In the limit $n\rightarrow\infty$ we have $(n-k)\rightarrow n$ and we conclude that

\begin{equation}
\rho_{C\infty}=\frac{R}{(T-R)}(k-1)^{-1}
\label{equiSzero}
\end{equation}

From Eq.(~\ref{equiSzero}) we find that for $(k-1) < \frac{R}{T-R}$ then $\rho_{c}->1$ whilst for $\infty > (k-1) > \frac{R}{T-R}$ we have coexistence of the two types. We see then that an increasingly randomized average degree can 'dilute' the advantage of the cooperative bond between parent and offspring with the onset of coexistence defined by the game payoffs. 
As is customary in the literature we can  make an identification between the language of Hamilton's principle  and the prisoner's dilemma payoffs  by letting $T=b$, $R=b-c$, $P=0$ and $S=-c$ . From this we conclude that cooperation will prevail if $k \leq b/c$. This result is in line with Hamilton's rule, frequent kin interactions promote cooperation, and its network extension ~\cite{LiebermanNowak} which states that networks of high connectance hamper cooperation as the average degree is an inverse measure of genetic relatedness.

Multiplying both sides of Eq.(~\ref{systemDepEquil}) by $k-1$ we get 

\begin{equation}
\rho_{C}(k-1)=\frac{n-k}{n}\frac{R}{T-R}=\overline{k_{c}}=constant
\label{systemDepKclinks}
\end{equation}
with the infinite version being:

\begin{equation}
\overline{k_{c\infty}}=\frac{R}{T-R}=constant
\label{Kclinks}
\end{equation}

Eqs. (~\ref{systemDepKclinks}) and (~\ref{Kclinks}) represent the expected number of cooperating links at fitness equilibrium an element will form due to randomly sampling the population. Since defectors can only acquire cooperator links through the random sampling we also have that 
\begin{equation}
\overline {k_{dc}}= \overline {k_{c}}=\frac{R}{T-R}
\label{kdc}
\end{equation}
which is the expected number of cooperators adjacent to a defector. 
Cooperators  also link to their parent so:

\begin{equation}
\overline{k_{cc}}=\overline{k_{dc}}+1= \frac{R}{T-R}+1=\frac{T}{T-R}
\label{kcc}
\end{equation}
for the expected number of cooperators adjacent to a cooperator. At equilibrium it should hold that $R\overline{k_{cc}}=T\overline{k_{dc}}$ so that $\overline k_{cc} = \frac{T}{R} \overline k_{dc}$.

\section{Numerical results}
\subsection{Co-existence properties and system size dependence}
\begin{figure}[ht]
	\centering
		\includegraphics[width=60mm]{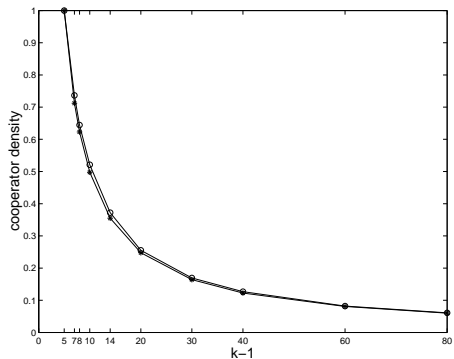}
	\caption{Equilibrium cooperator density as a function of $k-1 = \left[5,7,8,10,14,20,30,40,60,80 \right]$ in linear scale for systems with $n=2000$. Simulation results (circle) and Eq.(~\ref{systemDepEquil}) prediction(star).}
	\label{fig1}
\end{figure}

\begin{figure}[ht]
	\centering
		\includegraphics[width=60mm]{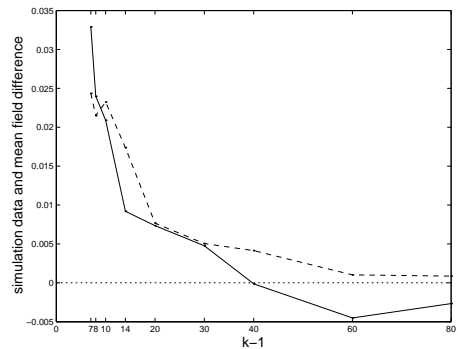}
	\caption{Difference between simulation results and the predictions of Eq.~\ref{systemDepEquil}. $n=2000$ dashed line and $n=1000$ solid line. The units on the y-axis are in percentage of cooperators in the population.}
	\label{fig2}
\end{figure}

In Fig.(~\ref{fig1}) we present results for $\rho_{c}$ as a function of $p_{o}n=k-1$ from agent based simulations of systems with $n=2000$. To compute $\rho_{c}$ we first average over 50 independent realizations with different initial conditions and then average again over the last 10 generations.
As mentioned in ~\cite{PollackC} a finite population undergoing stochastic replication with no mutation will eventually enter one of its absorbing states, which in our case is $\rho_{c}=1$ or $\rho_{c}=0$.The time to absorption however may be extremely long and all we can do is measure the observables of this pre-absorption transient. In what follows we will refer to the pre-abrorption transient states as equilibrium and focus on issues regarding the time to absorption in future work.
From Fig.(~\ref{fig1})  we can see that the simulation $\rho_{c}$ is roughly in good agreement with the values calculated from the mean field approximation Eq.(~\ref{systemDepEquil}).For $k-1 < 5$ there is no coexistence and the absorbing state is one with $\rho_{c}=1$, for $k-1 > 5$ we have coexistence that approximates a power law as $\rho_{c}\propto (k-1)^{\beta}$ with $\beta\approx-1$. Notice however from Fig.(~\ref{fig2}) how simulation data for $n=2000$ systems always give a slightly higher cooperator density for all degrees than Eq.(~\ref{systemDepEquil}) predicts,  whereas for the $n=1000$ systems the cooperator density is lower than expected for high average degrees. By comparing the two lines in Fig.(~\ref{fig2}) we can also see how increasing system size brings the simulation results closer and closer to the mean field predictions.

\begin{figure}[ht]
	\centering
		\includegraphics[width=60mm]{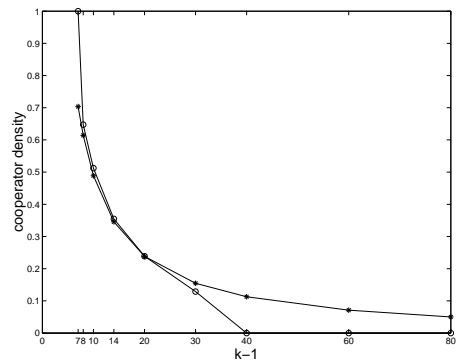}
	\caption{Equilibrium cooperator density as a function of $k-1 = \left[7,8,10,14,20,30,40,60,80 \right]$ in linear scale for systems with $n=400$. Simulation results (circle) and Eq.(~\ref{systemDepEquil}) prediction(star).}
	\label{fig3}
\end{figure}

\begin{figure}[ht]
	\centering
		\includegraphics[width=60mm]{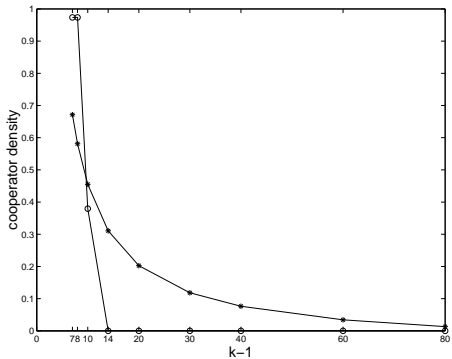}
	\caption{Equilibrium cooperator density as a function of $k-1 = \left[7,8,10,14,20,30,40,60,80 \right]$ in linear scale for systems with $n=100$. Simulation results (circle) and Eq.(~\ref{systemDepEquil}) prediction(star). For $k-1=7,8,10$ the $\rho_{c}$ values do not denote the average proportion of cooperators over the realizations but rather the frequency by which the state $\rho_{c}=1$ appears.}
	\label{fig4}
\end{figure}

This invites us then to look at the behaviour of systems of even smaller size. In Fig.(~\ref{fig3}) we show simulation data for systems of size $n=400$. As we can see these systems approach the behaviour of an infinite system for intermediate values of the average degree. For large(i.e. $k-1=40$) or small (i.e. $k-1=7$) average degree, where one of the two types has a large fitness advantage over the other, the systems will fall into the absorbing states $\rho_{c}=0$ or $\rho_{c}=1$ . For even smaller $n=100$ systems,Fig.(~\ref{fig4}),  the effect is more pronounced and co-existence is impossible. 


From the above we then conclude that our mean field approximation is unable to account for size dependent behaviour and can only provide a good approximation for large systems. Eq.(~\ref{systemDepEquil}) predicts that as we increase system size we should approach the infinite system $\rho_{c}$ from below. Our data suggests that whether we approach the infinite system $\rho_{c}$ from below or above depends on the average degree.
As a  first attemtpt to account for this behaviourwe hypothesize that in smaller systems density fluctuations bring forth the absorbing states $\rho_{c}=1$ and $\rho_{c}=0$. 

\begin{figure}[ht]
	\centering
		\includegraphics[width=60mm]{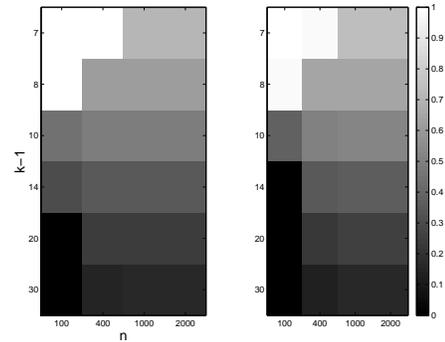}
	\caption{Iteration of replicator equation with gaussian fluctuations (left) and simulation data (right) for systems of different sizes and average degrees. Both sets of data are averages over 60 experiments. The cooperator density is mapped to a linear color index ranging from black ($\rho_{c}=0$) to white  ($\rho_{c}=1$).}
	\label{fig5}
\end{figure}

To this extent we perform  a numerical iteration of the discrete replicator, Eq.(~\ref{discReplicator}) with the fitnesses of the two types calulated by Eqs.(~\ref{wc}) and (~\ref{wd}). In each step after the next generation densities have been calculated we generate a Gaussian fluctuation with a mean of zero and a standard deviation of $\sigma$ which we then add to the cooperator density and remove from the defector density (we call this a stochastic replicator). To calculate $\sigma$ for a given $n$ we time average the standard deviation of $\rho_{c}$ over every generation and then ensemble average over our number of realizations. This data suggests that $\sigma\approx\frac{1}{4\sqrt{n}}$. In Fig.(~\ref{fig5}) we show a comparison of  simulation data and the numerical iteration of our stochastic replicator equation. Comparing the left and right image in the figure we can see how the stochastic replicator reproduces the monomorphic absorbing states for small $n$. Coexistence starts as the systems get larger and the average degree is at intermediate values. the fit between the output of the stochastic replicator and our simulation data is not perfect, for example for $n=100$ and $k-1=14$ the simulation data give $\rho_{c}=0$ while iteration of the replicator gives $\rho_{c}=0.308$. Sources of discrepancy between the stochastic replicator and our death/birth dynamics could be the generational versus sequential process and also the selection method, tournament versus fitness proportional (see ~\cite{PollackB}). This is to be determined in future work.

To construct an iterative stochastic replicator, that also explicitly incorporates the death/birth process, will also require an iterative model for the evolution of the degree distributions since there is a tight coupling between fitness and network structure. This brings us neatly to our next section.

\subsection{Degree distributions}

The degree distributions in our model are essentially also fitness distributions. Hence an analysis of their equilibrium and dynamical properties  captures all the fundamentals of the evolving dynamical system. In this paper we will briefly examine the distributions at the point where the strategies are in fitness equilibrium. Future work will concentrate on developing iterative coupled degree distribution and strategy models.
Let us here introduce the following notation:

\begin{itemize}
\item $p_{c}(k)$. The probability to find a cooperator with $k$ neighbours.
\item $p_{d}(k)$. The probability to find a cooperator with $k$ neighbours.
\item $p_{c}(k_{c})$. The probability to find a cooperator with $k_{c}$ cooperator neighbours.
\item $p_{d}(k_{c})$. The probability to find a defector with $k_{c}$ cooperator neighbours.
\item $p_{c}(k_{d})$. The probability to find a cooperator with $k_{d}$ defector neighbours.
\item $p_{d}(k_{d})$. The probability to find a defector with $k_{c}$ defector neighbours.
\end{itemize}

\begin{figure}[ht]
	\centering
		\includegraphics[width=60mm]{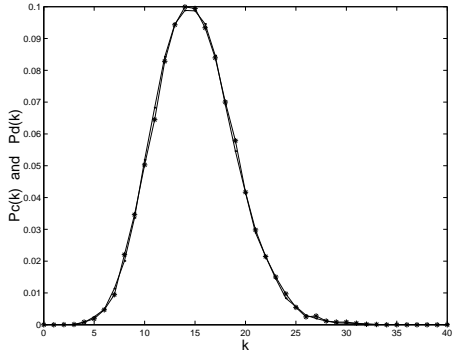}
	\caption{$p_{c}(k)$ (star), $p_{d}(k)$ (point) for systems with $N=2000$, $\overline{k}=15$ and steady state  $\rho_{c}=0.3718$. The data points are ensemble averages over 60 independent experiments. }
	\label{fig6}
\end{figure}

\begin{figure}[ht]
	\centering
		\includegraphics[width=60mm]{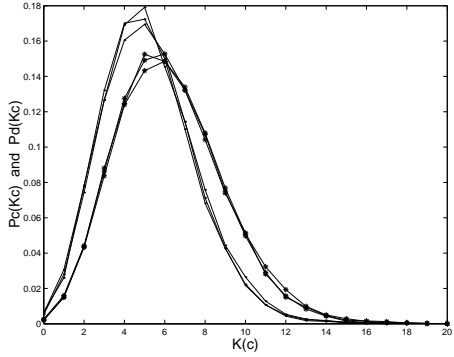}
	\caption{$p_{c}(k_{c})$ (star), $p_{d}(k_{c})$ (point) for three different systems with steady states characterised by $\rho_{c}=0.7355$, $\rho_{c}=0.5206$ and $\rho_{c}=0.3715$. The data points are ensemble averages over 60 independent experiments. }
	\label{fig7}
\end{figure}

In Fig.(~\ref{fig6}) we show plots from our individual based simulations for $p_{c}(k)$ and $p_{d}(k)$. As we can see there is nothing to distinguish the two types in terms of their neighbourhood size. In Fig.(~\ref{fig7}) we show plots of $p_{c}(k_{c})$ and $p_{d}(k_{c})$. We now see that $p_{c}(k_{c})$ and $p_{d}(k_{c})$ differ from each other. Furthermore their shape and relative position to each other remain invariant as $\rho_{c}$ varies. This invariance is a direct consequence of fitness equilibrium since the degree distributions are essentially fitness distributions. 
We can check that the distributions satisfy a coexisting population scenario by using the distribution data to calculate the following probabilities:
\begin{itemize}
\item $p_{n_{c},n_{c}+1}$, the probability that the number of cooperators will increase by one.
\item $p_{n_{c},n_{c}-1}=p_{n_{d},n_{d}+1}$ the probability that the defector population will increase by one
\item $p_{n_{c}n_{c}} =n_{n_{d},n_{d}}$ the probability that the population composition will remain the same.
\end{itemize}
These probabilities should satisfy:
\begin{equation}
p_{n_{c},n_{c}+1} + p_{n_{d},n_{d}+1} + p_{n_{c},n_{c}}=1
\label{eq:transProb}
\end{equation}

In our death/birth dynamics four events can occur. A cooperator can die with probability $d_{c}$ or be born with $ b_{c}$, and a defector can die with $d_{d}$ or be born with $ b_{d}$. This permits us to right the transition probabilities as:
\begin{itemize}
\item$p_{n_{c},n_{c}+1}= d_{d} \cdot  b_{c}$
\item$p_{n_{c},n_{c}-1}= d_{c} \cdot  b_{d}$
\item$p_{n_{c},n_{c}1}= d_{c} \cdot  b_{c}+d_{d} \cdot  b_{d}$
\end{itemize}
The death probabilities for the two types are simply proportional to their densities so that $d_{c}=\frac{n_{c}}{n}$ and $d_{d}=\frac{n_{d}}{n}$.

When we select two elements to test for reproduction the one with the highest total payoff reproduces with probability $1$. We will distinguish between the cases where the two elements are of the same type and when they are not. When the elements are of the same type then, since there is no mutation, an element of that type is added to the population. A cooperator will be born through this event with probability $\frac{n_{c}(n_{c}-1)}{n(n-1)}$ and a defector with probability $\frac{n_{d}(n_{d}-1)}{n(n-1)}$.

It is worthy to note that when we compare two individuals of the same type we could have just as well allowed for the less fit to reproduce.This is because when an individual is born all of its edges,except the one to the parent, are created at random and what is inherited is only the strategy. The situation becomes  different if the elements can inherit a portion of their links from their parent (i.e. $p_{e} > 0$). Then what is inherited is the type plus the 'environment' and it would matter whether we allowed the fitter or the less fit individual to reproduce. 

When we select two individuals of different type we further distinguish between the cases where the individuals have the same fitness and when they do not. When they have the same fitness then we select one at random to reproduce regardless of type. As mentioned previously  two elements of different type can have the same fitness only when $\frac{k_{cc}}{k_{dc}}=\frac{R}{T}$, we can then write the probability  of an individual being born by chance after been selected for the tournament together with another individual of opposite type but equal fitness as.

\begin{equation}
b_{eq}=\frac{1}{2}\sum_{i=0}p_{c}(\frac{T}{T-R}i)\frac{n_{c}}{n}\cdot p_{d}(\frac{R}{T-R}i)\frac{n_{d}}{n-1}
\end{equation}
with $i=0,1,2,...,int\left\{\frac{n(T-R)}{T}\right\}$ as before.

When we select a cooperator and a defector with different fitness values the individual with the highest fitness will reproduce. This depends on where on their respective degree distributions they sit in relation to each other.A cooperator with  $1 \leq k_{cc} \leq \frac{R}{T-R}i$ can win a defector if the defector has a $k_{dc}$ up to and including $k_{cc}-i$.  We can then write the probability that a cooperator was born because it won the tournament against a defector as:

\begin{equation}
b_{cw}=2\sum_{i}\sum_{k_{cc}} p_{c}(k_{cc})\frac{n_{c}}{n}\cdot \sum_{k_{dc}} p_{d}(k_{dc})\frac{n_{d}}{n-1}
\end{equation}

The limits are
\begin{itemize}
\item$i=0,1,2,...,int\left\{\frac{n(T-R)}{T}\right\}$
\item $\frac{T}{T-R}(i-1) \leq k_{cc} \leq \frac{T}{T-R}i-1$ 
\item $0 \leq k_{dc} \leq k_{cc}-i$.
\end{itemize}

For a defector via a similar payoff argument we get

\begin{equation}
P_{dw}=2\sum_{i}\sum_{k_{dc}} p_{d}(k_{dc})\frac{n_{d}}{n}\cdot \sum_{k_{cc}} p_{c}(k_{cc})\frac{n_{c}}{n-1}
\end{equation}

with limits
\begin{itemize}
\item$i=0,1,2,...,int\left\{\frac{n(T-R)}{T}\right\}$
\item $\frac{T}{T-R}(i-1) \leq k_{dc} \leq \frac{T}{T-R}i-1$ 
\item $0 \leq k_{cc} \leq k_{dc}+i-1$.
\end{itemize}

In table~\ref{table1} we have tabulated the birth and transition probabilities using the ensemble averaged $p_{c}(k_{c})$ and $p_{d}(k_{c})$ from simulation data for systems with $n=2000$. As is evident the transition probabilities between the two types are very close, also note how the birth probabilities of the two types are very close to their respective densities such that $\frac{\rho_{c}}{ \rho_{d}}=\frac{b_{c}}{b_{d}}$.We thus conclude that for this $n$ our systems are indeed in fitness equilibrium and well described by the mean field as in Eq.(~\ref{systemDepEquil}).

\begin{center}
\begin{table}[h!b!p!]
\begin{tabular*}{\columnwidth}{@{\extracolsep{\fill}}cccccc}
	\hline
		$k-1$ & \vline $\rho_{c}/\rho_{d}$ \vline & $ b_{c}/b_{d}$ \vline & $p_{n_{c},n_{c}+1}$ \vline & $p_{n_{d}->n_{d}+1}$ \vline & $p_{n_{c}->n_{c}}$\\
		\hline
		7 & 2.7822 & 2.7872  & 0.1938 & 0.1934 &  0.6128\\
		8 & 1.8121 & 1.8523  & 0.2232 &  0.2184 & 0.5584\\
		10 & 1.0846  &  1.0810   &    0.2479 & 0.2488 & 0.5033\\
		14 & 0.5913   & 0.5978    &  0.2340  &  0.2315   & 0.5345\\
		20 & 0.3419  &  0.3409   &   0.1888 &   0.1893 &   0.6219\\
		30 & 0.2024  &  0.2030    &  0.1401  &  0.1396  &  0.7203\\
		40 & 	0.1448  &  0.1472   &   0.1119  &  0.1101  &  0.7780\\
		60 & 0.0894  &  0.0926    & 0.0777   & 0.0751  &  0.8471\\
		\hline
\end{tabular*}
\label{table1}
\caption{densities of strategists together with birth and transition probabilities as a function of increasing average degree for systems with $N=2000$}
\end{table}
\end{center}

Finally let us mention that armed with expressions for the transition probabilities and a mean field set of equations that describe how the degree distributions change after a death/birth event we will be able to iterate an equation for the evolution of $\rho_{c}$ of the form:

\begin{equation}
\rho_{c}^{'} = p_{n_{c},n_{c}} \cdot \rho_{c} + p_{n_{c},n_{c}+1} \cdot (\rho_{c}+ \frac{1}{n}) + p_{n_{c},n_{c}-1} \cdot (\rho_{c}- \frac{1}{n})
\end{equation}

\subsection{The effect of strategy mutations}
If we introduce mutations, such that the offspring can have the opposite strategy from the parent with probability $p_{m}$ coexistence can persist indefinitely. The equilibrium size dependent $\rho_{c}$ then becomes:

\begin{equation}
\rho{c}=\frac{(n-k)R-p_{m}(n-1)(T+R)}{n(k-1)(T-R)}
\label{sizeDepMut}
\end{equation}

and as $n->\infty$

\begin{equation}
\rho{c}=\frac{R-p_{m}(T+R)}{(T-R)}
\label{NonSizeDepMut}
\end{equation}

\begin{figure}[ht]
	\centering
		\includegraphics[width=60mm]{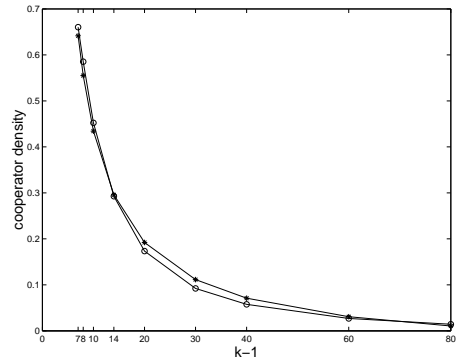}
	\caption{Equilibrium cooperator density as a function of $k-1 = \left[7,8,10,14,20,30,40,60,80 \right]$ in linear scale for systems with $n=100$. Simulation results (circle) and Eq.(~\ref{sizeDepMut}) prediction(star). There is a strategy mutation rate at birth of $p_{m}=0.01$. }
	\label{fig8}
\end{figure}

In Fig.(~\ref{fig8}) we show $\rho_{c}$ for systems with $n=100$ together with the mean field predictions of Eq.(~\ref{sizeDepMut}).

As we can see from the figure the introduction of $p_{m}$ helps suppress the fluctuations that in small systems can destroy co-existence completely (compare Figs.(~\ref{fig8}) and (~\ref{fig4})). Producing individuals of the opposite type has the obvious effect of avoiding the absorbing states.
Notice that a mutant offspring can only benefit from the link to the parent when the parent is a cooperator and the offspring a defector.
This suggests a mechanism for how the mutations might regulate the fluctuations in density that bring about the absorbing states. When there is a majority of cooperators (in low average degree $k$ environments) a mutant defector offspring will have a high fitness compared to the average in the population, thus will be more likely to reproduce. As defectors start to spread, and link to their parents, they again start to have lower average fitness  compared to defectors since in the low $k$ environment defectors cannot amass the number of cooperator neighbours they need to overcome the absence of payoff from the link to their parent.On the other hand when there is a defector majority a mutant cooperator offspring will most likely have the same fitness as its parent since $S=P=0$, thus cooperators can spread due to the parent-offspring link. When their numbers start to rise cooperators are at a disadvantage again as the high average degree enviroment spells their doom.

\section{Conclusions and future work}
To summarize we have showed how co-existing populations can persist in an ever changing dynamic environment. Our results are in line with Hamilton's rule of kin selection operating on a stochastically created network. Moreover we have briefly touched on the role of phenotypic mutation and its ability to regulate co-existence in smaller populations. We plan to continue our research in the role of fluctuations, the influence of selection methods and creating a model for the evolution of the degree distributions. Perhaps more importantly we plan to examine systems with $p_{e}>0$. In this case the offspring inherits a strategy and a set of succesfull relationships so that selection acts as to preserve succesfull network units and not just individual behaviours.

The authors kindly thank Chris Rhodes for help with proofreading.


\end{document}